\documentclass[aps,twocolumn,showpacs,prl,amsmath,amssymb,floatfix,superscriptaddress]{revtex4-1}
\usepackage{amsmath,amssymb,graphicx,color}
\usepackage{bm}
\usepackage{bbm}
\usepackage[caption=false]{subfig}
\usepackage[usenames,dvipsnames]{xcolor}
\usepackage[colorlinks=true,citecolor=Cerulean,linkcolor=RubineRed,urlcolor=Cerulean]{hyperref}
\usepackage[normalem]{ulem}
\usepackage{soul,xcolor}

\newcommand{\avg}[1]{\ensuremath{\langle #1 \rangle}}

\usepackage{fixltx2e,amsmath}
\MakeRobust{\eqref}

\usepackage{mathptmx} 

\begin{document}
\setstcolor{blue}
\title{Finite-Time Thermodynamics of Fluctuations in Microscopic Heat Engines}

\author{Gentaro Watanabe} 
\affiliation{Department of Physics and Zhejiang Institute of Modern Physics, Zhejiang University, Hangzhou, Zhejiang 310027, China}
\affiliation{Zhejiang Province Key Laboratory of Quantum Technology and Device, Zhejiang University, Hangzhou, Zhejiang 310027, China}
\author{Yuki Minami}
\affiliation{Department of Physics and Zhejiang Institute of Modern Physics, Zhejiang University, Hangzhou, Zhejiang 310027, China}

\date{\today}

\begin{abstract}
Fluctuations of thermodynamic quantities become non-negligible and play an important role when the system size is small. We develop finite-time thermodynamics of fluctuations in microscopic heat engines whose environmental temperature and mechanical parameter are driven periodically in time. Within the slow-driving regime, this formalism universally holds in a coarse-grained time scale whose resolution is much longer than the correlation time of the fluctuations, and is shown to be consistent with the relation analogous to the fluctuation-dissipation relation. Employing a geometric argument, a scenario to simultaneously minimize both the average and fluctuation of the dissipation in the Carnot cycle is identified. For this simultaneous optimization, the existence of a zero eigenvalue of the singular metric for the scale invariant equilibrium state is found to be essential. Furthermore, we demonstrate that our optimized protocol can improve the dissipation and its fluctuation over the current experiment.
\end{abstract}

\maketitle

\textit{Introduction.---} Technological development has enabled us to fabricate heat engines using small systems \cite{Hugel02,Steeneken11,Blickle12,Quinto-Su14,Martinez16,Serra-Garcia16,Martinez17,Klaers17,Argun17}. To understand the performance of such small engines, theory beyond the conventional thermodynamics for macroscopic systems, thermodynamics of small systems, is called for. Recent synergy between technology and the remarkable success of stochastic thermodynamics \cite{Sekimoto98,Sekimotobook10,Jarzynski11,Seifert12,VandenBroeck14,Ciliberto17,Nicolis17} in understanding thermodynamic properties of small systems \cite{Bustamante05} including micromachines using active matter \cite{Krishnamurthy16,Pietzonka19,Fodor21} and biological systems \cite{Toyabe15} has led to a surge of activity on the study of microscopic heat engines \cite{Sekimoto00,Schmiedl08,Dechant15,Brandner15,Dechant17,Brandner20,Strasberg21}, where effects of thermal fluctuation can play an important role \cite{Jop08,Martinez15, Hoppenau13,Holubec14,Rana14,Cerino15,Holubec17,Holubec18,Dechant19,carnotfluct,Saryal21,Mohanta21,Holubec21,  Verley14a,Verley14b,Proesmans15a,Proesmans15b,Park16,Saha18,Manikandan19,Vroylandt20,  Pal17,Sinitsyn11,  Proesmans17,Pietzonka18,Barato18,Koyuk19a,Koyuk19b,Timpanaro19,Kamijima21,Xu21}.

For reversible heat engines under quasistatic operations, thermodynamics provides the celebrated Carnot bound on their efficiency, and some universal relations are also known even for fluctuations \cite{carnotfluct}. However, in the practical situations, heat engines are operated within a finite period of cycle. There, losses due to dissipation is inevitable, and how to reduce such dissipative loss is a crucial issue. Furthermore, since the amount of the dissipative loss in microscopic heat engines significantly varies for each trial (e.g., the fluctuation of the efficiency is of order unity in the experiment of Ref.~\cite{Martinez16}), how to suppress the fluctuation of the dissipation is an equally important agenda, which stands out as an open problem. Here, we formulate finite-time thermodynamics of fluctuations in microscopic heat engines which is applicable to the slow-driving regime and coarse-grained timescale. This formalism provides us a geometric description for fluctuation of the dissipation under finite-time operations. Geometric descriptions \cite{Weinhold75,Ruppeiner79,Salamon84,Gilmore84,Schlogl85,Brody94,Ruppeiner95} of the mean value of the dissipative loss are known for both the macroscopic \cite{Salamon83,Andresen84,Nulton85} and microscopic systems \cite{Crooks07,Sivak12,Zulkowski12,Brandner20,Vu21}. It is noted that such a geometric description is also possible for its fluctuation. We apply our framework to the current experiment and provide an optimum scheme to minimize both the mean and the fluctuation of the dissipation simultaneously unlike trade-off optimization (see also \cite{Miller20} for trade-off optimization between work fluctuation and efficiency and \cite{Plata19,Plata20a,Plata20b,Frim21} for recent works on optimization of the mean value). Interestingly, it is found that the scale invariance of the equilibrium state is essential for such a simultaneous optimization, and this optimization is possible only for cycles solely consisting of isothermal and isentropic strokes as the Carnot cycle. Since the dissipation is closely related to the efficiency, which is also fluctuating in microscopic heat engines, the reduction of its mean and fluctuation leads to an enhancement and stabilization of the efficiency.

\textit{Setup.---}
Consider a classical microscopic heat engine whose working substance is always in contact with an environment whose temperature $T$ is controllable. This is a standard setting for the experiments of the classical microscopic heat engines \cite{Blickle12,Martinez16,Martinez17,Argun17}. The Hamiltonian $H_{\lambda_w}$ of the working substance has an external mechanical parameter $\lambda_w$ (or $V$). Therefore, the system has two control parameters $\lambda_w$ and $\lambda_u \equiv T$, and the engine protocol is specified by a closed path $\mathcal{C}$ in the parameter space spanned by the vector $\lambda_\mu \equiv (\lambda_w,\, \lambda_u)$. The parameters are varied cyclically to drive the state (more precisely, the phase space distribution function $p$) of the working substance to be periodic in time with the period $\tau$ of the cycle.

The generalized forces $(X_w,\, X_u)$ conjugate to the parameters $(\lambda_w,\, \lambda_u)$ are $(X_w,\, X_u) \equiv (P,\, S) = (-\partial H_{\lambda_w}/\partial \lambda_w, -\ln{p})$, where $P\equiv -\partial H_{\lambda_w}/\partial \lambda_w$ is the generalized pressure and $S \equiv - \ln{p}$ is the stochastic entropy. Note that $X_w$ and $X_u$ are random variables as functions of the phase space point $\Gamma$.

The loss of the energy due to driving with nonzero speed is quantified by the {\it dissipated availability} $A$ introduced by Salamon and Berry \cite{Salamon83}: $A \equiv U - W$ with $W$ being the work output by the engine and $U$ being the effective energy input from the environment (not to be confused with the internal energy) defined respectively as
\begin{align}
  W \equiv& \oint_\mathcal{C} P\, dV = \int_0^\tau dt \left(- \frac{\partial H_{\lambda_w}}{\partial \lambda_w} \right) \dot{\lambda}_w = \int_0^\tau dt\, X_w \dot{\lambda}_w\,,\label{eq:w}\\
  U \equiv& \oint_\mathcal{C} T\, dS = \int_0^\tau dt\, T \frac{d}{dt}(- \ln{p}) = \int_0^\tau dt\, \lambda_u \dot{X}_u\,,\label{eq:u}
\end{align}
where the dot denotes the time derivative. Here, $A$, $W$, and $U$ are random variables. Note that $T$ in $U$ is the temperature of the environment instead of that of the working substance, and $U$ is different from heat input in general, but agrees with the latter in the quasistatic limit.

In the following, we discuss the fluctuation of random variables. Fluctuation $\Delta Y$ of a random variable $Y$ is given by $\Delta Y \equiv Y - \avg{Y}$, where $\avg{\cdots}$ means an ensemble average.

Since the working substance is always in contact with the thermal environment, correlations of the physical quantities decay exponentially in time provided the driving is slow enough. In the coarse-grained time scale whose resolution is much longer compared to the decay time, such an exponential decay function can be well approximated by a delta function multiplied by its decay time constant (or the correlation time). Therefore, in the slow-driving regime and coarse-grained time scale, the two-time correlation function of the fluctuations could universally be written in the form of
\begin{align}
  \avg{\Delta X_\mu(t)\, \Delta X_\nu(t')} = 2 \avg{\Delta X_\mu(t)\, \Delta X_\nu(t)} \tau_{\mu\nu}(t)\, \delta(t-t')\,,\label{eq:2tcf}
\end{align}
where $\tau_{\mu\nu}(t)$ $[=\tau_{\nu\mu}(t)]$ is the correlation time between $\Delta X_\mu$ and $\Delta X_\nu$ at time $t$. In the following, we will formulate finite-time thermodynamics of fluctuations of the dissipated availability based on Eq.~(\ref{eq:2tcf}). The ansatz (\ref{eq:2tcf}) is the minimal prescription to introduce the time scale to thermodynamics through $\tau_{\mu\nu}$ which depends on microscopic details of the system.

First, we consider the variance $\avg{\Delta A^2}$ of $A$ for a cycle given by $\avg{\Delta A^2} = \avg{\Delta U^2} + \avg{\Delta W^2} - 2 \avg{\Delta U\, \Delta W}$. Regarding the variance of $W$, using Eq.~(\ref{eq:2tcf}) and the condition of the closed cycle such that the phase space distribution function at $t=0$ and $\tau$ are the same, $p(0) = p(\tau)$, we obtain
\begin{align}
  \avg{\Delta W^2} =& \int_0^\tau dt \int_0^\tau dt'\, \left[ \avg{X_w(t) X_w(t')} - \avg{X_w(t)} \avg{X_w(t')} \right]\,\nonumber\\
  & \times \dot{\lambda}_w(t)\, \dot{\lambda}_w(t')\nonumber\\
  =&\, 2 \int_0^\tau dt\, \avg{\Delta X_w^2(t)}\, \tau_{ww}(t)\, \dot{\lambda}_w^2(t)\,.\label{eq:varw}
\end{align}
By integrating by parts, $U$ given by Eq.~(\ref{eq:u}) can be rewritten as $U = -\int_0^\tau X_u(t)\, \dot{\lambda}_u(t)\, dt + [X_u(\tau)-X_u(0)]\, \lambda_u(0)$ since $\lambda_u(\tau) = \lambda_u(0)$ but $X_u(\tau) \ne X_u(0)$ in general. Thus, using Eq.~(\ref{eq:2tcf}) and the closed cycle condition as in the way to get Eq.~(\ref{eq:varw}), we obtain
\begin{align}
  \avg{\Delta U^2} =&\, 2 \int_0^\tau dt\, \avg{\Delta X_u^2(t)}\, \tau_{uu}(t)\, \dot{\lambda}_u^2(t) + 2\, \avg{\Delta X_u^2(0)}\, \lambda_u^2(0)\,,\label{eq:varu}\\
  \avg{\Delta W \Delta U} =&\, -2 \int_0^\tau dt\, \avg{\Delta X_w(t)\, \Delta X_u(t)}\, \tau_{wu}(t)\, \dot{\lambda}_w(t)\, \dot{\lambda}_u(t)\,.\label{eq:covwu}
\end{align}
From Eqs.~(\ref{eq:varw}), (\ref{eq:varu}), and (\ref{eq:covwu}), we finally obtain
\begin{align}
  \avg{\Delta A^2} = \int_0^\tau dt\, g^{(2)}_{\mu\nu}(t)\, \dot{\lambda}_\mu(t)\, \dot{\lambda}_\nu(t) + 2\, \avg{\Delta X_u^2(0)}\, \lambda_u^2(0)\label{eq:vara}
\end{align}
with
\begin{align}
  g^{(2)}_{\mu\nu}(t) \equiv 2\, \tau_{\mu\nu}(t)\, \avg{\Delta X_\mu(t) \Delta X_\nu(t)} \equiv 2\, \tau_{\mu\nu}(t)\, \sigma_{\mu\nu}(t)\,
\end{align}
for each $(\mu, \nu)$ element, where $\sigma_{\mu\nu} \equiv \avg{\Delta X_\mu(t)\, \Delta X_\nu(t)}$ is the covariance matrix element between $X_\mu$ and $X_\nu$. Summation should be taken over $(w, u)$ components for the repeated $\mu$ and $\nu$ in Eq.~(\ref{eq:vara}) and hereafter. For a continuous operation taking over consecutive $N$ cycles, an average $\overline{\avg{\Delta A^2}}$ of the fluctuation of $A$ per cycle in the limit of $N \rightarrow \infty$ is given by
\begin{align}
  \overline{\avg{\Delta A^2}} =& \lim_{N \rightarrow \infty} \frac{1}{N} \int_0^{N\tau} dt\, g^{(2)}_{\mu\nu}(t) \dot{\lambda}_\mu(t)\, \dot{\lambda}_\nu(t)\nonumber\\
  =& \int_0^{\tau} dt\, g^{(2)}_{\mu\nu}(t)\, \dot{\lambda}_\mu(t)\, \dot{\lambda}_\nu(t)\,,\label{eq:barvara}
\end{align}
since the second term of $\avg{\Delta A^2}$ in Eq.~(\ref{eq:vara}) comes solely from the end points and their contribution vanishes by taking the average over an infinite number of cycles.

Since the covariance matrix $\sigma_{\mu\nu}$ is positive definite, $g^{(2)}_{\mu\nu}$ is also positive definite when $\tau_{ww}\tau_{uu} = \tau_{wu}^2$. Therefore, as long as $\tau_{ww}\tau_{uu} = \tau_{wu}^2$ is satisfied, $g^{(2)}_{\mu\nu}$ can be regarded as a metric tensor. Note that the resulting expression of the variance $\overline{\avg{\Delta A^2}}$ given by Eq.~(\ref{eq:barvara}) is in the similar form as the average $\avg{A} = \int_0^\tau dt\, g^{(1)}_{\mu\nu}(t)\, \dot{\lambda}_\mu(t)\, \dot{\lambda}_\nu(t)$ derived in Ref.~\cite{Brandner20}, but with a different metric tensor $g^{(2)}_{\mu\nu}$. From the Cauchy-Schwarz inequality, $[\int dt\, f(t)\, g(t)]^2 \le [\int dt\, f^2(t)]\, [\int dt\, g^2(t)]$ with $f(t) = (g^{(2)}_{\mu\nu} \dot{\lambda}_\mu \dot{\lambda}_\nu)^{1/2}$ and $g(t) = 1$, we get
\begin{align}
  \overline{\avg{\Delta A^2}} \ge \frac{(\mathcal{L}^{(2)})^2}{\tau}\label{eq:boundvara}
\end{align}
with the thermodynamic length $\mathcal{L}^{(2)}$ for the variance of $A$:
\begin{align}
  \mathcal{L}^{(2)} \equiv \int_0^\tau dt\, \sqrt{g^{(2)}_{\mu\nu}(t) \dot{\lambda}_\mu(t)\, \dot{\lambda}_\nu(t)} = \oint \sqrt{g^{(2)}_{\mu\nu}\, d\lambda_\mu\, d\lambda_\nu}\,.\label{eq:length2}
\end{align}
The equality in Eq.~(\ref{eq:boundvara}) holds if and only if $g^{(2)}_{\mu\nu}(t) \dot{\lambda}_\mu(t) \dot{\lambda}_\nu(t)$ is constant or identically equal to zero.

\textit{Relation between $g^{(1)}_{\mu\nu}$ and $g^{(2)}_{\mu\nu}$.---}
Next, we discuss the relation between $g^{(1)}_{\mu\nu}$ and $g^{(2)}_{\mu\nu}$. Here, we assume that the phase space distribution function $p$ follows the Fokker-Planck equation (or more generally, the Kramers-Moyal equation):
\begin{align}
  \frac{\partial p(\Gamma,\, t)}{\partial t} = L(\Gamma,\, t)\, p(\Gamma,\, t)
\end{align}
with $L(\Gamma,\, t)$ being the derivative operator (so-called the Fokker-Planck operator or the Kramers-Moyal operator), so that the equilibrium distribution $p^{\rm eq}$ satisfies the detailed balance condition:
\begin{align}
  L(\Gamma,\, t)\, p^{\rm eq}(\Gamma) \cdots = p^{\rm eq}(\Gamma)\, L^\dagger(\sigma\Gamma,\, t) \cdots\,,
\end{align}
where $\dagger$ denotes the adjoint and $\sigma$ ($= +1$ or $-1$) is the symmetry factor of the phase space variable under the time reversal operation. For sufficiently slow driving such that its time scale $\lambda_\mu/\dot{\lambda}_\mu$ is much larger than the relaxation time of the working substance, we write
\begin{align}
  p(\Gamma, t) = [1+\xi(\Gamma, t)]\, p^{\rm eq}(\Gamma; t)\,,
\end{align}
where $p^{\rm eq}(\Gamma; t) \equiv \exp{[-\beta(t)\, H_{\lambda_w}(\Gamma; t)]}/Z_t$ is the equilibrium state for the instantaneous parameter values $(\lambda_w(t),\, \lambda_u(t))$ at $t$, $\beta(t) \equiv 1/T(t)$ is the instantaneous inverse temperature, $Z_t \equiv \int d\Gamma\, \exp{[-\beta(t)\, H_{\lambda_w}(\Gamma; t)]}$ is the partition function, and $\xi \ll 1$ describes the deviation from $p^{\rm eq}$. Within the linear approximation with respect to the small quantities $\dot{\lambda}_\mu$ and $\xi$, we obtain (see \cite{Supplement} for details)
\begin{align}
  \xi(\Gamma, t) \simeq -\beta(t) \int_0^\infty ds\, e^{L^\dagger(\sigma\Gamma,\, t)\, s}\, \left[ X_\mu(t) - \avg{X_\mu(t)}_{t,\, {\rm eq}} \right]\, \dot{\lambda}_\mu(t)\,\label{eq:xi}
\end{align}
with $\avg{\cdots}_{t,\, {\rm eq}} \equiv \int d\Gamma\, p^{\rm eq}(\Gamma; t)\cdots$ being the average for the instantaneous equilibrium state at $t$. Within the linear response regime, the metric tensor $g^{(1)}_{\mu\nu}$ for $\avg{A}$ can be written as $g^{(1)}_{\mu\nu}(t) \equiv - 2^{-1} [R_{\mu\nu}(t) + R_{\nu\mu}(t)]$ with the response coefficients $R_{\mu\nu}$ defined by $\avg{X_\mu(t)} = \int d\Gamma\, X_\mu(t)\, (1+\xi) p^{\rm eq} \equiv \avg{X_\mu(t)}_{t,\, {\rm eq}} + R_{\mu\nu}(t)\, \dot{\lambda}_{\nu}(t)$ \cite{Brandner20}. Using Eq.~(\ref{eq:xi}), it can be shown that the ansatz (\ref{eq:2tcf}) leads to the relation between the two metric tensors $g^{(1)}_{\mu\nu}$ and $g^{(2)}_{\mu\nu}$ as
\begin{align}
  g^{(2)}_{\mu\nu}(t) = 2\, T(t)\, g^{(1)}_{\mu\nu}(t)\,,\label{eq:fdr}
\end{align}
which is analogous to the fluctuation-dissipation relation.

\textit{Application to the Brownian Carnot cycle.---}
An overdamped Brownian particle in a harmonic oscillator trapping potential is a typical setup commonly employed in experiments of microscopic heat engines \cite{Blickle12,Martinez16,Seifert12,Martinez17,Argun17}. Taking this setup as an example, we shall apply our framework to evaluate the fluctuation of the performance of the microscopic heat engine. For simplicity, we consider a one-dimensional case, and thus the phase space point $\Gamma$ of the overdamped system can be specified solely by the position of the Brownian particle $q$: $\Gamma = \{q\}$. Suppose, with the mechanical control parameter $\lambda_w$, the external one-dimensional harmonic oscillator trap $V_{\lambda_w}(q)$ is given by
\begin{align}
  V_{\lambda_w}(q) = \frac{\lambda_w}{2} q^2\,.\label{eq:v}
\end{align}
This system can be described by the time-dependent Ornstein-Uhlenbeck process whose Fokker-Planck equation is \cite{Risken_book,Gardiner_book}
\begin{align}
  \frac{\partial}{\partial t} p(q,\, t) = \frac{\lambda_w(t)}{\gamma} \frac{\partial}{\partial q}\left[q\, p(q,\, t)\right] + \frac{D(t)}{\gamma} \frac{\partial^2}{\partial q^2} p(q,\, t)\,,\label{eq:fpeq}
\end{align}
where $D$ is the diffusion constant and $\gamma$ is the friction coefficient.

Now, we consider a cycle whose driving speed $\dot{\lambda}_\mu$ is finite, but sufficiently slow (i.e., the time scale $\lambda_\mu/\dot{\lambda}_\mu$ is much larger than the relaxation time) so that the working substance is always close to the instantaneous equilibrium state and
\begin{align}
  D(t) = \beta^{-1}(t)\,.
\end{align}
Thus, to obtain $\avg{A}$ and $\overline{\avg{\Delta A^2}}$ up to the linear order of small quantities, $g^{(i)}_{\mu\nu}$ in their expressions can be evaluated for the instantaneous equilibrium state since it is already multiplied by a factor of the linear order of small quantities: $\int_0^\tau dt \dot{\lambda}_\mu \dot{\lambda}_\nu$.

Since the Hamiltonian of the overdamped system is given solely by the potential energy, i.e. $H_{\lambda_w} = V_{\lambda_w}$, the instantaneous equilibrium state is $p^{\rm eq} = \exp{[-\beta \lambda_w q^2/2]}/Z_t$ with $Z_t = \sqrt{2\pi T / \lambda_w}$. For the equilibrium state, each element of the covariance matrix $\sigma_{\mu\nu} = \avg{\Delta X_\mu \Delta X_\nu}_{t,\, {\rm eq}}$ reads $\sigma_{ww} = (T/\lambda_w)^2/2$, $\sigma_{wu} = \sigma_{uw} = - T/2\lambda_w$, and $\sigma_{uu} = 1/2$.

Next, we consider the correlation times $\tau_{\mu\nu}$ in the equilibrium state. The correlation functions $\avg{\Delta X_\mu(t)\, \Delta X_\nu(0)}_{\rm eq}$ in the equilibrium state read $\avg{\Delta X_w(t)\, \Delta X_w(0)}_{\rm eq} = 4^{-1} C_{q^2}(t)$, $\avg{\Delta X_w(t)\, \Delta X_u(0)}_{\rm eq} = \avg{\Delta X_u(t)\, \Delta X_w(0)}_{\rm eq} = - (\lambda_w/4T) C_{q^2}(t)$, and $\avg{\Delta X_u(t)\, \Delta X_u(0)}_{\rm eq} = (\lambda_w/2T)^2 C_{q^2}(t)$ with $C_{q^2}(t) \equiv \avg{q^2(t)\, q^2(0)}_{\rm eq} - \avg{q^2}_{\rm eq}^2$. Since the time-dependent part of these correlation functions $\avg{\Delta X_\mu(t)\, \Delta X_\nu(0)}_{\rm eq}$ is given by $C_{q^2}(t)$, all the correlation times should be equal: $\tau_{ww}=\tau_{uu}=\tau_{wu}=\tau_{uw}$. Therefore, they satisfy $\tau_{ww} \tau_{uu} = \tau_{wu}^2$, so that $g^{(2)}_{\mu\nu}$ is positive definite and $g^{(i)}_{\mu\nu}$ ($i=1, 2$) can indeed be regarded as metric tensors.
Using the nonstationary solution $p(q,t|q', t')$ (with $t\ge t'$) of the Fokker-Planck equation (\ref{eq:fpeq}) for constant $\lambda_w$ and $D=\beta^{-1}$ at their instantaneous values \cite{Risken_book} and the stationary solution $p_s=p^{\rm eq}$ obtained for $t-t' \gg \tau_{\mu\nu}$, we get $\avg{q^2(t) q^2(0)}_{\rm eq} = \iint dq_1 dq_2\, q_1^2 q_2^2\, p(q_1,t|q_2,0) p_s(q_2) = (D/\lambda_w)^2 (1+ 2e^{-2\lambda_w t/ \gamma})$ and $\avg{q^2}_{\rm eq} = \int dq\, q^2 p_s(q) = D/\lambda_w$, and thus $C_{q^2}(t) = 2(D/\lambda_w)^2 e^{-2\lambda_w t/ \gamma}$. Therefore, we finally obtain
\begin{align}
  \tau_{ww} = \tau_{uu} = \tau_{wu} = \frac{\gamma}{2\lambda_w}\,.\label{eq:taumunu}
\end{align}

\begin{figure}[t!]
\centering
\includegraphics[width=0.93 \columnwidth]{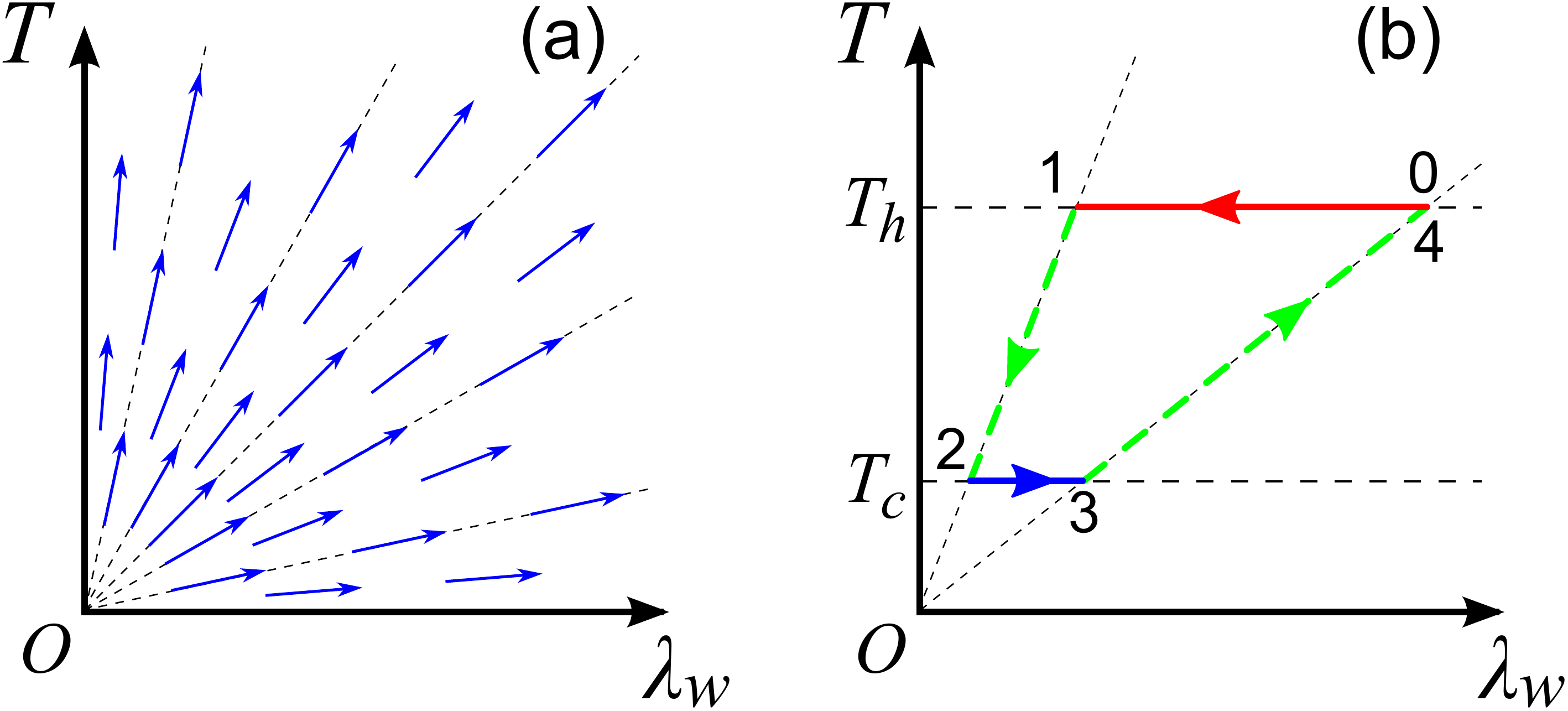}
\caption{(a) Eigenvector field of $g^{(2)}_{\mu\nu}$ corresponding to the zero eigenvalue shown in the $T$-$\lambda_w$ plane. The dotted straight lines are a guide for the eye showing the directions of the zero eigenvalue of $g^{(2)}_{\mu\nu}$ for different values of $T/\lambda_w$. (b) Brownian Carnot cycle, which consists of two isothermal strokes at $T_h$ ($0 \rightarrow 1$) and $T_c$ ($2 \rightarrow 3$), and two isentropic strokes ($1 \rightarrow 2$ and $3 \rightarrow 4$).
}
\label{fig:tlambda}
\end{figure}

From $\sigma_{\mu\nu}$ evaluated for the instantaneous equilibrium state and Eq.~(\ref{eq:taumunu}), the metric tensor $g^{(2)}_{\mu\nu}$ reads
\begin{equation}
  g^{(2)}_{\mu\nu} = \frac{\gamma}{2\lambda_w}
  \begin{bmatrix}
    (T/\lambda_w)^2 & -T/\lambda_w \\
    -T/\lambda_w & 1 \\
  \end{bmatrix}\,.\label{eq:g2}
\end{equation}
Remarkably, this metric is singular with a zero eigenvalue whose corresponding eigenvector (normalized to unity) is $[(\lambda_w/T)^2 +1]^{-1/2}\, (\lambda_w/T, 1)^\intercal$. Thus, the direction of the zero eigenvalue of $g^{(2)}_{\mu\nu}$ is $dT/d\lambda_w = T/\lambda_w$, which is along a straight line on the $T$-$\lambda_w$ plane connecting each point at $(\lambda_w, T)$ and the origin [see Fig.~\ref{fig:tlambda}(a)]. Since $g^{(1)}_{\mu\nu} \propto g^{(2)}_{\mu\nu}$, $g^{(1)}_{\mu\nu}$ also has a zero eigenvalue and the above argument applies to $g^{(1)}_{\mu\nu}$ as well. The other eigenvalue of $g^{(2)}_{\mu\nu}$ is $\gamma (2\lambda_w)^{-1} [1+(T/\lambda_w)^2]$ and the corresponding eigenvector is orthogonal to the one for the zero eigenvalue, i.e., $\propto (-1, \lambda_w/T)^\intercal$.

It is noted that, along the path of the zero eigenvalue of $g^{(i)}_{\mu\nu}$ (i.e., a straight line on the $T$-$\lambda_w$ plane for a given value of $T/\lambda_w$), the mean value $\avg{S}_{\rm eq}$ of the stochastic entropy is constant:
\begin{align}
  \avg{S}_{\rm eq} = - \avg{\ln{p^{\rm eq}}}_{\rm eq} = 2^{-1} \left[ 1 + \ln{2\pi} - \ln{(\lambda_w/T)} \right]\,.
\end{align}
Therefore, the isentropic process defined as the one that conserves $\avg{S}_{\rm eq}$ \cite{Martinez15,Martinez16} is along the path of the zero eigenvalue of $g^{(i)}_{\mu\nu}$.

Now we consider what is called the Brownian Carnot cycle (hereafter, we refer to it as the Carnot cycle for simplicity) realized in an experiment by Mart\'inez {\it et al.} \cite{Martinez16}, which consists of two isothermal strokes at hot ($T_h$) and cold ($T_c$) temperatures, and two isentropic strokes \cite{note:isentropic} as shown in Fig.~\ref{fig:tlambda}(b). The duration of the hot ($0 \rightarrow 1$) and cold ($2 \rightarrow 3$) isothermal strokes is denoted by $\tau_h$ and $\tau_c$, and that of the isentropic strokes $1 \rightarrow 2$ and $3 \rightarrow 4$ is denoted by $\tau_{1 \rightarrow 2}$ and $\tau_{3 \rightarrow 4}$, respectively. Thus the period $\tau$ of the cycle is given by $\tau = \tau_{\rm isoth} + \tau_{\rm isen}$ with $\tau_{\rm isoth} \equiv \tau_h + \tau_c$ and $\tau_{\rm isen} \equiv \tau_{1 \rightarrow 2} + \tau_{3 \rightarrow 4}$ being the total duration of the isothermal strokes and isentropic strokes in a cycle, respectively.

Suppose the duration of each stroke in the cycle is given and it is large enough so that the linear approximation is valid. Now, we shall optimize the protocol of each stroke under its given duration to minimize $\avg{A}$ and $\overline{\avg{\Delta A^2}}$. First of all, it is noted that, since the isentropic strokes are along the path of the zero eigenvalue of $g^{(1)}_{\mu\nu}$ and $g^{(2)}_{\mu\nu}$, these strokes have no contribution to $\avg{A}$, $\overline{\avg{\Delta A^2}}$, and the thermodynamic lengths $\mathcal{L}^{(i)} \equiv \displaystyle \oint \sqrt{g^{(i)}_{\mu\nu}\, d\lambda_\mu d\lambda_\nu}$ of the cycle \cite{note:fluct}. Therefore, the isentropic strokes are completely irrelevant to this optimization, and what we have to do is to optimize only the isothermal strokes to fulfill the lower bounds of the inequalities:
\begin{align}
  \avg{A} = \avg{A_h} + \avg{A_c} \ge \frac{(\mathcal{L}^{(1)}_{h})^2}{\tau_{h}} + \frac{(\mathcal{L}^{(1)}_{c})^2}{\tau_{c}}\label{eq:bound1}
\end{align}
\cite{Sekimoto97,Brandner20} and
\begin{align}
  \overline{\avg{\Delta A^2}} = \overline{\avg{\Delta A_h^2}} + \overline{\avg{\Delta A_c^2}} \ge \frac{(\mathcal{L}^{(2)}_{h})^2}{\tau_{h}} + \frac{(\mathcal{L}^{(2)}_{c})^2}{\tau_{c}}\,,\label{eq:bound2}
\end{align}
where $A_h$ and $A_c$ are the dissipated availability in the hot and cold isothermal strokes, and $\mathcal{L}^{(i)}_h \equiv \displaystyle\int_{0 \rightarrow 1} \sqrt{g^{(i)}_{\mu\nu} d\lambda_\mu d\lambda_\nu}$ and $\mathcal{L}^{(i)}_c \equiv \displaystyle\int_{2 \rightarrow 3} \sqrt{g^{(i)}_{\mu\nu} d\lambda_\mu d\lambda_\nu}$ are the thermodynamic length of the path of the hot and cold isothermal strokes, respectively. Writing $\lambda_{w,i}$ for $\lambda_w$ at node $i$, from Eqs.~(\ref{eq:g2}) and (\ref{eq:fdr}) we obtain $\mathcal{L}_h^{(i)} = \sqrt{2^{i-1} \gamma T_h^i}\, (\lambda_{w,1}^{-1/2} - \lambda_{w,0}^{-1/2})$ ($i=1$, $2$). Since $T_h/\lambda_{w,1} = T_c/\lambda_{w,2}$ and $T_h/\lambda_{w,0} = T_c/\lambda_{w,3}$, we further get $\mathcal{L}_c^{(i)} = (T_c/T_h)^{(i-1)/2} \mathcal{L}_h^{(i)}$ ($i=1$, $2$). As has been discussed for Eq.~(\ref{eq:boundvara}), the equalities of Eqs.~(\ref{eq:bound1}) and (\ref{eq:bound2}) hold when $g^{(i)}_{\mu\nu}(t)\, \dot{\lambda}_\mu(t)\, \dot{\lambda}_\nu(t)$ is constant in time, or equivalently, when we sweep the parameter $\lambda_\mu$ in a way such that the time interval to traverse the path element $d\lambda_\mu$ is proportional to the thermodynamic length $d\mathcal{L}^{(i)}$ of this path element. Since $g^{(i)}_{\mu\nu}\, \dot{\lambda}_\mu\, \dot{\lambda}_\nu = g^{(i)}_{ww}\, \dot{\lambda}_w^2$ in the isothermal strokes, the above condition is satisfied by sweeping as
\begin{align}
  |\dot{\lambda}_w| \propto \lambda_w^{3/2}\,,\label{eq:optimal}
\end{align}
and we get $\avg{A} = (\tau_h^{-1} + \tau_c^{-1}) (\mathcal{L}_h^{(1)})^2$ and $\overline{\avg{\Delta A^2}} = [\tau_h^{-1} + (T_c/T_h) \tau_c^{-1}] (\mathcal{L}_h^{(2)})^2$.
It is noted that, since $g^{(1)}_{\mu\nu}$ and $g^{(2)}_{\mu\nu}$ are related through Eq.~(\ref{eq:fdr}) and $T$ is constant in each isothermal stroke, the optimal protocol (\ref{eq:optimal}) applies to both $\avg{A}$ and $\overline{\avg{\Delta A^2}}$. Namely, once the duration of each stroke is given, both $\avg{A}$ and $\overline{\avg{\Delta A^2}}$ for the Carnot cycle are optimized simultaneously, which is in contrast to the trade-off optimization between, e.g., work fluctuation and efficiency \cite{Miller20}.

As can be seen from the above discussion, such simultaneous minimization of $\avg{A}$ and $\overline{\avg{\Delta A^2}}$ by the same protocol within finite time is possible only for cycles solely consisting of isothermal strokes and isentropic strokes. First of all, isentropic strokes have zero contribution to $\avg{A}$ and $\overline{\avg{\Delta A^2}}$ because of the zero eigenvalue of the metric in our system. In addition, only isothermal strokes are nontrivial strokes with nonzero $\avg{A}$ and $\overline{\avg{\Delta A^2}}$ that allow us to minimize both $\avg{A}$ and $\overline{\avg{\Delta A^2}}$ simultaneously by the same protocol $\lambda_\mu(t)$. Among cycles whose number of strokes $N_s$ is less than or equal to $5$, the Carnot cycle is the only possibility to form a meaningful cycle with nonzero work output solely with isothermal and isentropic strokes \cite{note:trivialstrokes}. Therefore, when the duration of each stroke is given, the Carnot cycle is the only case with $N_s \le 5$ which allows us to simultaneously minimize $\avg{A}$ and $\overline{\avg{\Delta A^2}}$.

It is noted that the existence of the zero eigenvalue is essential for the above simultaneous minimization, and such a singular metric is due to the scale invariance of the equilibrium state. In the overdamped case, once the trapping potential is scale invariant, i.e. $V_{\lambda_w}$ satisfies $V_{\lambda_w}(a q) = |a|^\alpha V_{\lambda_w}(q)$ for some $\alpha$ with $a$ being the scaling factor, so is the system itself, and as a consequence, the metric becomes singular. A power-law function is a representative example of the scale invariant function, and the above simultaneous minimization of $\avg{A}$ and $\overline{\avg{\Delta A^2}}$ is possible also for a general power-law trapping potential $V_{\lambda_w}(q) \propto \lambda_w\, |q|^k$ with positive $k$ \cite{Supplement}.

If only the total duration of the isothermal strokes $\tau_{\rm isoth}$ is given instead of $\tau_h$ and $\tau_c$ separately, we can further optimize to reduce either $\avg{A}$ or $\overline{\avg{\Delta A^2}}$, respectively. This optimization holds irrespective of the duration $\tau_{\rm isen}$ and the protocol of the isentropic strokes, provided the driving in these strokes is slow enough. From the condition that $g^{(i)}_{\mu\nu}(t)\, \dot{\lambda}_\mu(t)\, \dot{\lambda}_\nu(t) = g_{ww}^{(i)} \dot{\lambda}_w^2$ is constant at the same value for both the two isothermal strokes, the optimal protocol is $|\dot{\lambda}_w| \propto \lambda_w^{3/2} T^{-i+1}$ ($i=1$, $2$), which yields $\tau_h=\tau_c=\tau_{\rm isoth}/2$ to optimize $\avg{A}$ and $\tau_h/\sqrt{T_h} = \tau_c/\sqrt{T_c}$ [i.e., $\tau_h = \tau_{\rm isoth}/(1+\sqrt{T_c/T_h})$ and $\tau_c = \tau_{\rm isoth} - \tau_h$] to optimize $\overline{\avg{\Delta A^2}}$ with their minimum vales $\avg{A} = (\mathcal{L}^{(1)})^2/\tau_{\rm isoth}$ and $\overline{\avg{\Delta A}^2} = (\mathcal{L}^{(2)})^2 /\tau_{\rm isoth}$, respectively.

Finally, we discuss how much the optimized protocol given by Eq.~(\ref{eq:optimal}) improves the dissipated availability and its fluctuation of the Carnot cycle compared to the one in the current experiment \cite{Martinez16}. In this experiment, $\lambda_{w}(t)$ in Eq.~(\ref{eq:v}) is controlled by the following protocol: $\lambda_{w}(t) = f(t)$ for $0 \leq t \leq \tau/2$ and $\lambda_{w}(t) = f(\tau -t)$ for $\tau/2 \leq t \leq \tau$ with $f(t)=4(\lambda_{w, 2}-\lambda_{w,0})(t/\tau)^2+\lambda_{w, 0}$. The experimental parameter values are as follows: $T_c=300$~K, $T_h=525$~K, $\lambda_{w,0}=20.0$~pN$\cdot \mu$m$^{-1}$, $\lambda_{w,2}=2.0$~pN$\cdot \mu$m$^{-1}$, the cycle period $\tau = 200$~ms, and the duration of each stroke is $\tau_{0 \rightarrow 1} = 0.26 \tau$, $\tau_{1 \rightarrow 2} = 0.24 \tau$, $\tau_{2 \rightarrow 3} = 0.25 \tau$, and $\tau_{3 \rightarrow 4} = 0.25 \tau$. In addition, the friction coefficient is estimated as $\gamma=8.4$~pN$\cdot \mu$m$^{-1}\cdot$ms~\cite{Supplement}. From these parameter values, the ratio of the dissipated availabilities for the experimental protocol and the optimized protocol given by Eq.~(\ref{eq:optimal}) and that for their fluctuations are calculated to be~\cite{Supplement}
\begin{align}
\frac{\langle A_{\rm opt} \rangle}{\langle A_{\rm exp} \rangle}=0.65, \quad  \frac{\,\,\,\overline{\langle \Delta A^2_{\rm opt} \rangle}\,\,\,}{\overline{\langle \Delta A^2_{\rm exp} \rangle}}=0.70.
\end{align}
Here, the subscripts ``opt'' and ``exp'' represent quantities obtained by the optimized and the experimental protocols, respectively. The optimized protocol improves the dissipated availability by $35$\% and its fluctuation by $30$\%. This means that the above optimization improves not only the efficiency $\epsilon$, which is given by $\epsilon \equiv \avg{W}/\avg{U} \simeq 1 - \avg{A}/\mathcal{W}$ with $\mathcal{W}$ $(=\avg{\mathcal{W}})$ being the work output by the quasistatic cycle \cite{Brandner20}, but also the stability of the fluctuating efficiency $\mathcal{E} \equiv W/U \simeq 1 - A/\mathcal{W}$ characterized by $\overline{\avg{\Delta \mathcal{E}^2}} \simeq \overline{\avg{\Delta A^2}}/\mathcal{W}^2$. Since the fluctuation of the stochastic efficiency $\mathcal{E}$ is nonnegligible in the Brownian Carnot engine (in the experiment of \cite{Martinez16}, the variance of $\mathcal{E}$ is of order unity, and the distribution function of $\mathcal{E}$ spreads even to the negative side), it is crucially important to reduce $\overline{\avg{\Delta A^2}}$ as well as $\avg{A}$.

\textit{Concluding remarks.---}
In summary, we have formulated finite-time thermodynamics of fluctuations in microscopic heat engines, which universally holds for the slow-driving regime and the coarse-grained time scale. This formalism provides a geometric description of the fluctuation of the dissipation whose metric is found to be consistent with the relation analogous to the fluctuation-dissipation relation. Applying our framework to the Carnot cycle, it has been found that both the average and the fluctuation of the dissipation can be minimized simultaneously when the duration of each stroke is given. Interestingly, we have seen that the scale invariance of the equilibrium state is essential for this simultaneous minimization. The benefit of this optimized protocol has been demonstrated for the current experiment \cite{Martinez16}.

Our framework should find broad applications to design not only energy efficient but also stable microscopic thermal machines. While we have considered classical systems, developing a corresponding formalism for quantum systems constitutes an important and challenging future problem. Further analysis based on Riemannian geometry for systems with a non-singular metric, such as a classical or quantum two-level system, etc., would lead to a deeper understanding on thermodynamics of fluctuations. For example, geometric interpretation of thermodynamics of fluctuations provided in our work may lead to another type of universal relations and bounds on fluctuations of the performance of microscopic heat engines distinct from thermodynamic uncertainty relations \cite{Barato15,Gingrich16,Macieszczak18,Seifert19,Falasco20}.

\bigskip
\begin{acknowledgments} 
This work was supported by NSF of China (Grants No.~11975199 and No.~11674283), by the Zhejiang Provincial Natural Science Foundation Key Project (Grant No.~LZ19A050001), and by the Zhejiang University 100 Plan.
\end{acknowledgments}

\clearpage
\onecolumngrid

\begin{center}
	
	\newcommand{\beginsupplement}{%
		\setcounter{table}{0}
		\renewcommand{\thetable}{S\arabic{table}}%
		\setcounter{figure}{0}
		\renewcommand{\thefigure}{S\arabic{figure}}%
	}
	
        \textbf{\large --- Supplemental Material --- \\Finite-Time Thermodynamics of Fluctuations in Microscopic Heat Engines}
\end{center}
\newcommand{\beginsupplement}{%
	\setcounter{table}{0}
	\renewcommand{\thetable}{S\arabic{table}}%
	\setcounter{figure}{0}
	\renewcommand{\thefigure}{S\arabic{figure}}%
}

\setcounter{equation}{0}
\setcounter{figure}{0}
\setcounter{table}{0}
\setcounter{page}{1}
\makeatletter
\renewcommand{\theequation}{S\arabic{equation}}
\renewcommand{\thefigure}{S\arabic{figure}}
\renewcommand{\bibnumfmt}[1]{[S#1]}
\renewcommand{\citenumfont}[1]{S#1}
\vspace{0.8 in}

\newcommand{\ve}[1]{\bm{#1}}


\vspace{-2 cm}

\subsection{Fluctuation-dissipation relation between the thermodynamic metrics}

For a cycle of small heat engines with period $\tau$ considered in the main paper, provided the variation of the parameters $\lambda_\mu(t)$ (with $\mu = w$, $u$) is sufficiently slow, the average $\avg{A_1}$ and the variance $\avg{\Delta A_1^2}$ of the dissipated availability $A_1$ for one cycle can be written using a metric tensor $g^{(i)}_{\mu\nu}$ as (The subscript ``$1$'' in $A_1$ shows that the quantity is obtained for a single cycle. Hereafter, the subscript ``$1$'' will be omitted, and the dissipated availability $A$ without a subscript means that the one obtained for one cycle.)
\begin{align}
  \avg{A} =& \int_0^\tau dt\, g^{(1)}_{\mu\nu}(t)\, \dot{\lambda}_\mu(t)\, \dot{\lambda}_\nu(t)\,,\label{eq:avga_suppl}\\
  \avg{\Delta A^2} =& \int_0^\tau dt\, g^{(2)}_{\mu\nu}(t)\, \dot{\lambda}_\mu(t)\, \dot{\lambda}_\nu(t) + 2 \avg{\Delta X_u^2(0)} \lambda_u^2(0)\,,\label{eq:vara_suppl}
\end{align}
with
\begin{align}
  g^{(2)}_{\mu\nu}(t) = 2\tau_{\mu\nu}(t)\, \avg{\Delta X_\mu(t)\, \Delta X_\nu(t)} \equiv 2 \tau_{\mu\nu}(t)\, \sigma_{\mu\nu}(t)\,.
\end{align}
Here, $\Delta X_\mu(t) \equiv X_\mu(t) - \avg{X_\mu(t)}$ is the fluctuation of the generalized force $X_\mu(t)$, $\tau_{\mu\nu}(t)$ is the correlation time between $\Delta X_\mu$ and $\Delta X_\nu$ at time $t$, and $\sigma_{\mu\nu}(t) \equiv \avg{\Delta X_\mu(t)\, \Delta X_\nu(t)}$ is the covariance matrix of $X_\mu$ and $X_\nu$ at $t$.

For some quantity $O_N$ obtained through a continuous operation taking over consecutive $N$ cycles, an average $\overline{O}$ of $O_N$ per cycle in the limit of $N \rightarrow \infty$ is defined as
\begin{align}
  \overline{O} \equiv \lim_{N \rightarrow \infty} \frac{1}{N} O_N\,.
\end{align}
Since the second term of $\avg{\Delta A^2}$ in Eq.~(\ref{eq:vara_suppl}) comes solely from the initial state, this contribution vanishes by taking the average over an infinite number of cycles:
\begin{align}
  \overline{\avg{\Delta A^2}} \equiv \lim_{N \rightarrow \infty} \frac{1}{N} \avg{\Delta A_N^2} = \lim_{N \rightarrow \infty} \frac{1}{N} \int_0^{N\tau} dt\, g^{(2)}_{\mu\nu}(t) \dot{\lambda}_\mu(t)\, \dot{\lambda}_\nu(t) = \int_0^{\tau} dt\, g^{(2)}_{\mu\nu}(t) \dot{\lambda}_\mu(t)\, \dot{\lambda}_\nu(t)\,.
\end{align}
In the last equality, we have used the fact that the system is cyclic: $\lambda_\mu(t+\tau) = \lambda_\mu(t)$ and $g^{(2)}_{\mu\nu}(t+\tau) = g^{(2)}_{\mu\nu}(t)$ for any $t$.

In the following, we show that there is a relation between the metrics $g^{(1)}_{\mu\nu}$ and $g^{(2)}_{\mu\nu}$ analogous to the fluctuation-dissipation relation (FDR) as
\begin{align}
  g^{(2)}_{\mu\nu}(t) = 2 k_BT(t)\, g^{(1)}_{\mu\nu}(t)\,\label{eq:fdr_suppl}
\end{align}
within the linear approximation with respect to the deviation from the instantaneous equilibrium state. Here, $k_B$ is the Boltzmann constant and $T(t)$ is the temperature at $t$.

To show the relation (\ref{eq:fdr_suppl}), we set the following assumptions.
\begin{itemize}
\item The time evolution of the phase space distribution function $p$ is governed by the Fokker-Planck equation, or more generally, by the Kramers-Moyal equation:
\begin{align}
  \frac{\partial p(\Gamma, t)}{\partial t} = L(\Gamma, t) p(\Gamma, t)\,,\label{eq:fp_suppl}
\end{align}
where $\Gamma$ is the phase space point and $L$ is the derivative operator (so-called the Fokker-Planck operator or the Kramers-Moyal operator).

\item Detailed balance condition (see, e.g., \cite{Risken_book_suppl}), which automatically holds for the Fokker-Planck and the Kramers-Moyal operators:
\begin{align}
  L(\Gamma, t)\, p^{\rm eq}(\Gamma)\,\,\, \cdots = p^{\rm eq}(\Gamma)\, L^\dagger(\sigma \Gamma, t)\,\,\, \cdots\,,\label{eq:db_suppl}
\end{align}
where $p^{\rm eq}(\Gamma)$ is the equilibrium phase space distribution function, $\sigma$ is the symmetry factor of the phase space variable under the time reversal operation (e.g., $\sigma =+1$ for the position and $\sigma = -1$ for the momentum), and the superscript ``$\dagger$'' denotes the adjoint: suppose $A$ is a real operator, then $A^\dagger$ is an adjoint operator of $A$ defined as $\int \phi A \psi\, d\Gamma = \int (A^\dagger \phi) \psi\, d\Gamma$ for any real functions $\phi$ and $\psi$. Note that Eq.~(\ref{eq:db_suppl}) is an operator equation which is valid when it is applied to an arbitrary function [i.e., the parts denoted by ``$\cdots$'' in Eq.~(\ref{eq:db_suppl})].

\item Slow driving such that the timescale $\lambda_\mu(t)/\dot{\lambda}_\mu(t)$ of changing the parameter $\lambda_\mu(t)$ is much larger than the relaxation time of the working substance to the equilibrium state so that the deviation $\xi$ from the equilibrium state $ p^{\rm eq}(\Gamma; t)$ for the instantaneous parameter values $\lambda_\mu(t)$ is small:
\begin{align}
  p(\Gamma, t) = \left[1 + \xi(\Gamma, t)\right] p^{\rm eq}(\Gamma; t)\label{eq:p_suppl}
\end{align}
with $\xi(\Gamma, t) \ll 1$. Here, $p^{\rm eq}(\Gamma; t)$ is the canonical distribution for the instantaneous mechanical parameter $\lambda_w(t)$ and the temperature $\lambda_u(t)=T(t)$ at time $t$ given by
\begin{align}
  p^{\rm eq}(\Gamma; t) \equiv \frac{e^{-\beta(t) H(\Gamma;\, t)}}{Z_t}\,,
\end{align}
where $\beta(t) \equiv 1/k_BT(t)$ is the inverse temperature, $H(\Gamma; t)$ is the Hamiltonian of the working substance, and $Z_t \equiv \int d\Gamma\, \exp[{-\beta(t)\, H(\Gamma; t)]}$ is the partition function.

\end{itemize}

In the following, we shall show that our formulation is consistent with the FDR-type relation given by Eq.~(\ref{eq:fdr_suppl}).

Using Eq.~(\ref{eq:p_suppl}) and within the linear approximation, the lhs of Eq.~(\ref{eq:fp_suppl}) can be written as
\begin{align}
  \partial_t p(\Gamma, t) &= \partial_t [(1+\xi)\, p^{\rm eq}(\Gamma; t)]\nonumber\\
  &\simeq (\partial_t \xi) p^{\rm eq} + \partial_t p^{\rm eq}\nonumber\\
  &\simeq \dot{\lambda}_\mu \frac{\partial p^{\rm eq}}{\partial \lambda_\mu}\,.\label{eq:dpdt_suppl}
\end{align}
Note that the first term in the second line of Eq.~(\ref{eq:dpdt_suppl}) is negligible since $\partial_t\xi$ is the second order of the small quantities. Then, $\dot{\lambda}_\mu \partial p^{\rm eq}/\partial \lambda_\mu$ for each component $\mu=w$ and $u$ can be calculated as
\begin{align}
  \dot{\lambda}_w(t)\, \frac{\partial}{\partial \lambda_w} p^{\rm eq} = \dot{\lambda}_w(t)\, \beta(t) \left[ X_w(t) - \avg{X_w(t)}_{t,\, {\rm eq}} \right] p^{\rm eq}(\Gamma; t)\,,\label{eq:dlambdaw_suppl}
\end{align}
and
\begin{align}
  \dot{\lambda}_u(t)\, \frac{\partial}{\partial \lambda_u} p^{\rm eq}  = \dot{\lambda}_u(t)\, \beta(t) \left[ -k_B \ln{p^{\rm eq}} - \avg{-k_B \ln{p^{\rm eq}} }_{t,\, {\rm eq}} \right] p^{\rm eq}(\Gamma; t)\,,\label{eq:dlambdau_suppl}
\end{align}
where $\avg{\cdots}_{t,\, {\rm eq}} \equiv \int d\Gamma\, p^{\rm eq}(\Gamma;\, t) \cdots$ is an average for the instantaneous equilibrium distribution at $t$.
Since $\dot{\lambda}_\mu$ is a small quantity, $p^{\rm eq}$ in $\ln{p^{\rm eq}}$ in the rhs of Eq.~(\ref{eq:dlambdau_suppl}) can be replaced by $p$ within the linear approximation:
\begin{align}
  \dot{\lambda}_u(t)\, \frac{\partial}{\partial \lambda_u} p^{\rm eq}
  &\simeq \dot{\lambda}_u(t)\, \beta(t) \left[ -k_B \ln{p} - \avg{-k_B \ln{p}}_{t,\, {\rm eq}} \right]\, p^{\rm eq}(\Gamma; t)\nonumber\\
  &= \dot{\lambda}_u(t)\, \beta(t) \left[X_u(t) - \avg{X_u(t)}_{t,\, {\rm eq}}\right]\, p^{\rm eq}(\Gamma; t)\,,
\end{align}
where $X_u(t) \equiv S(t) = -k_B \ln{p(\Gamma, t)}$ is the stochastic entropy. Therefore, the lhs of Eq.~(\ref{eq:fp_suppl}) can finally be written as
\begin{align}
  \frac{\partial }{\partial t} p(\Gamma, t) \simeq \frac{p^{\rm eq}(\Gamma;\, t)}{k_BT(t)} \left[ X_\mu(t) - \avg{X_\mu(t)}_{t,\, {\rm eq}} \right] \dot{\lambda}_\mu(t)\,.\label{eq:fplhs_suppl}
\end{align}

Next, we consider the rhs of Eq.~(\ref{eq:fp_suppl}). Using Eq.~(\ref{eq:p_suppl}) and the detailed balance condition (\ref{eq:db_suppl}), the rhs of Eq.~(\ref{eq:fp_suppl}) can be rewritten as
\begin{align}
  L(\Gamma, t)\, p(\Gamma, t) &= L(\Gamma, t)\, p^{\rm eq}(\Gamma; t)\, \left[1+\xi(\Gamma, t)\right]\nonumber\\
  &= p^{\rm eq}(\Gamma; t)\, L^\dagger(\sigma\Gamma, t)\, \left[1+\xi(\Gamma, t) \right]\nonumber\\
  &= p^{\rm eq}(\Gamma; t)\, L^\dagger(\sigma\Gamma, t)\, \xi(\Gamma, t)\,.\label{eq:fprhs_suppl}
\end{align}

Substituting Eqs.~(\ref{eq:fplhs_suppl}) and (\ref{eq:fprhs_suppl}) into Eq.~(\ref{eq:fp_suppl}), we obtain
\begin{align}
  L^\dagger(\sigma\Gamma, t)\, \xi(\Gamma, t) \simeq \beta(t) \left[ X_\mu(t) - \avg{X_\mu(t)}_{t,\, {\rm eq}} \right]\, \dot{\lambda}_\mu(t)\,.\label{eq:fp2_suppl}
\end{align}
The formal solution of Eq.~(\ref{eq:fp2_suppl}) is
\begin{align}
  \xi(\Gamma, t) \simeq -\beta(t) \int_0^\infty ds\, e^{L^\dagger(\sigma\Gamma,\, t)\, s}\, \left[ X_\mu(t) - \avg{X_\mu(t)}_{t,\, {\rm eq}} \right]\, \dot{\lambda}_\mu(t)\,.
\end{align}
Thus, the ensemble average of $X_\mu(t)$ for the distribution $p$ is given by
\begin{align}
  \avg{X_\mu(t)} \simeq \int d\Gamma\, X_\mu(t) \left[ 1+\xi(\Gamma, t) \right]\, p^{\rm eq}(\Gamma; t) \equiv \avg{X_\mu(t)}_{t,\, {\rm eq}} + R_{\mu\nu}(t)\, \dot{\lambda}_\nu(t)\label{eq:xmuavg_suppl}
\end{align}
with
\begin{align}
  R_{\mu\nu}(t) &\simeq -\beta(t) \int_0^\infty ds\, \int d\Gamma\, \left[ e^{L^\dagger(\Gamma,\, t)\, s}\, \Delta X_\mu(t) \right]\, \Delta X_\nu(t) p^{\rm eq}(\Gamma; t)\nonumber\\
  &= -\beta(t) \int_0^\infty ds\, \avg{\Delta X_\mu(t-s)\, \Delta X_\nu(t)}_{t,\,\, {\rm eq}}\nonumber\\
  &\simeq -\beta(t) \int_0^\infty ds\, \avg{\Delta X_\mu(t-s)\, \Delta X_\nu(t)}_t\,,\label{eq:rmunu_suppl}
\end{align}
where $\avg{\cdots}_t \equiv \int d\Gamma\, p(\Gamma, t) \cdots$. In obtaining the first line of Eq.~(\ref{eq:rmunu_suppl}), we have rewritten as $X_\mu p^{\rm eq} e^{L^\dagger(\sigma\Gamma,\, t)\, s} \cdots = X_\mu e^{L(\Gamma,\, t)\, s} p^{\rm eq} \cdots = [e^{L^\dagger(\Gamma,\, t)\, s} X_\mu] p^{\rm eq} \cdots$ using the detailed balance condition and the adjoint relation. It is also noted that the average for $p^{\rm eq}(\Gamma; t)$ in $R_{\mu\nu}$ can be replaced by that for $p(\Gamma, t)$ (and vice versa) within the linear approximation since $R_{\mu\nu}$ in Eq.~(\ref{eq:xmuavg_suppl}) is multiplied by a small quantity $\dot{\lambda}_\nu(t)$.

In the coarse-grained timescale, we have
\begin{align}
  \avg{\Delta X_\mu(t)\, \Delta X_\nu(t')} = 2\tau_{\mu\nu}(t)\, \avg{\Delta X_\mu(t)\, \Delta X_\nu(t)}\, \delta(t-t')\,.\label{eq:dxmudxnu_suppl}
\end{align}
Thus, $R_{\mu\nu}$ in this timescale reads
\begin{align}
  R_{\mu\nu}(t) \simeq -\beta(t)\, \tau_{\mu\nu}(t)\, \avg{\Delta X_\mu(t) \Delta X_\nu(t)}\,.
\end{align}
Since $g^{(1)}_{\mu\nu}$ is given by \cite{Brandner20suppl}
\begin{align}
  g^{(1)}_{\mu\nu}(t) = -\frac{1}{2} \left[ R_{\mu\nu}(t) + R_{\nu\mu}(t)\right]\,,
\end{align}
$g^{(1)}_{\mu\nu}$ in the coarse-grained timescale reads
\begin{align}
  g^{(1)}_{\mu\nu} \simeq \frac{\tau_{\mu\nu}(t)}{k_BT(t)} \avg{\Delta X_\mu(t)\, \Delta X_\nu(t)} \equiv \frac{\tau_{\mu\nu}(t)}{k_BT(t)} \sigma_{\mu\nu}(t)
\end{align}
with the covariance matrix $\sigma_{\mu\nu}(t) \equiv \avg{\Delta X_\mu(t)\, \Delta X_\nu(t)}$. Therefore, our relation (\ref{eq:dxmudxnu_suppl}) for the coarse-grained timescale yields
\begin{align}
  g^{(2)}_{\mu\nu}(t) \simeq 2k_BT(t)\, g^{(1)}_{\mu\nu}(t)\nonumber\,,
\end{align}
which is an analogous to the FDR.

\subsection{Scale invariant potential}

Here we shall show that the metric for an overdamped Brownian particle trapped in a scale invariant potential is singular and the direction of its zero eigenvalue is along an isentrope. For concreteness, we specifically consider a scale invariant external potential of the following power-law form:
\begin{align}
  V_{\lambda_w}(q) = \frac{\lambda_w}{k} |q|^k\label{eq:vsuppl}
\end{align}
with $k > 0$, but it is also possible to show that the metric is singular for a general scale invariant potential satisfying $V_{\lambda_w}(aq) = |a|^\alpha\, V_{\lambda_w}(q)$ for some $\alpha$ with $a$ being the scaling factor (see below). The harmonic oscillator potential discussed in the main paper is given by Eq.~(\ref{eq:vsuppl}) for $k=2$.

The equilibrium state $p^{\rm eq}$ for the instantaneous values of $\lambda_w$ and the inverse temperature $\beta$ is
\begin{align}
  p^{\rm eq} = Z_t^{-1} \exp{\bigl[-\beta \lambda_w |q|^k / k\bigr]}
\end{align}
with
\begin{align}
  Z_t = \int_{-\infty}^{\infty} dq\, \exp{\bigl[-\beta \lambda_w |q|^k / k\bigr]} = \frac{2 \Gamma(k^{-1})}{k (\beta \lambda_w/k)^{1/k}}\,,
\end{align}
where $\Gamma(x) \equiv \int_0^{\infty} z^{x-1} e^{-z} dz$ is the Gamma function. For the instantaneous equilibrium state, each element of the covariance matrix $\sigma_{\mu\nu} = \avg{\Delta X_\mu\, \Delta X_\nu}_{t,\, {\rm eq}}$ reads $\sigma_{ww} = (k_B T/\lambda_w)^2/k$, $\sigma_{wu} = \sigma_{uw} = -k_B^2 T/k\lambda_w$, and $\sigma_{uu} = k_B^2/k$.

The correlation functions $\avg{\Delta X_\mu(t)\, \Delta X_\nu(0)}_{\rm eq}$ in the equilibrium state read $\avg{\Delta X_w(t)\, \Delta X_w(0)}_{\rm eq} = k^{-2} C_{|q|^k}(t)$, $\avg{\Delta X_w(t)\, \Delta X_u(0)}_{\rm eq} = \avg{\Delta X_u(t)\, \Delta X_w(0)}_{\rm eq} = -(\lambda_w/kT) C_{|q|^k}(t)$, and $\avg{\Delta X_u(t)\, \Delta X_u(0)}_{\rm eq} = (\lambda_w/kT)^2 C_{|q|^k}(t)$ with $C_{|q|^k}(t) \equiv \avg{|q(t)|^k\, |q(0)|^k}_{\rm eq} - \avg{|q|^k}_{\rm eq}^2$. Similarly to the discussion in the main paper for the harmonic oscillator potential, since all the time dependence of these correlation functions is in $C_{|q|^k}(t)$, all the correlation time should be equal: $\tau_{\rm corr} \equiv \tau_{ww} = \tau_{uu} = \tau_{wu} = \tau_{uw}$.

Therefore, the metric tensor $g_{\mu\nu}^{(2)}$ reads
\begin{align}
  g_{\mu\nu}^{(2)} = \frac{2 k_B^2 \tau_{\rm corr}}{k}
  \begin{bmatrix}
  (T/\lambda_w)^2 & -T/\lambda_w \\
  -T/\lambda_w & 1
  \end{bmatrix}.\label{eq:gsuppl}
\end{align}
Obviously, the determinant of this metric is zero, so that it is singular with a zero eigenvalue. The eigenvector (normalized to unity) corresponding to the zero eigenvalue is $[(\lambda_w/T)^2 +1]^{-1/2}\, (\lambda_w/T,\, 1)^\intercal$, and thus its direction is $dT/d\lambda_w = T/\lambda_w$, which is along a straight line on the $T$-$\lambda_w$ plane connecting each point at $(\lambda_w,\, T)$ and the origin as in the case of harmonic oscillator potential discussed in the main paper. For a general scale invariant potential satisfying $V_{\lambda_w}(aq) = |a|^\alpha\, V_{\lambda_w}(q)$, its metric can also be shown to be singular. This can be understood from the fact that the phase-space integral in the covariance matrix can be evaluated as, e.g., $\int_{-\infty}^\infty dq\, e^{-\beta\, V_{\lambda_w}(q)} = \int_{-\infty}^\infty dq\, e^{-\beta\, V_{\lambda_w}(a q)}|_{a=1} = \int_{-\infty}^\infty da\, e^{-\beta\, V_{\lambda_w}(a q)}|_{q=1} = \int_{-\infty}^\infty da\, e^{-\beta\, |a|^\alpha\, V_{\lambda_w}(1)}$, which is essentially equivalent to the case of the power-law potential with an exponent $\alpha$.

The mean value of the stochastic entropy for $V_{\lambda_w}$ of Eq.~(\ref{eq:vsuppl}) is given by
\begin{align}
  \avg{S}_{\rm eq} = - k_B \avg{\ln{p^{\rm eq}}}_{\rm eq} = k^{-1}\, k_B\, \bigl[ 1 - \ln{(\beta\lambda_w/k)} \bigr] + k_B\, \ln{[2\Gamma(k^{-1})/k]}\,.
\end{align}
Therefore, an isentrope is given by $\lambda_w/T = \mbox{const.}$, which is along the direction of the zero eigenvalue of the metric.

\subsection{ Dissipated availability of the Brownian Carnot engine with the optimized and the current experiment protocols }

We estimate how much the optimized protocol given by Eq.~(25) in the main paper improves the dissipated availability of the Brownian Carnot engine compared to the current experimental protocol~\cite{Martinez16suppl}. In the experiment, $\lambda_{w}(t)$ of the harmonic potential [Eq.~(17) in the main paper] is controlled by the following protocol:
\begin{align}
\lambda_{w}(t) = \left\{
\begin{array}{ll}
f(t) &   \quad \mbox{for} \quad 0 \leq t \leq \tau/2, \\
f(\tau -t) &\quad  \mbox{for} \quad  \tau/2 \leq t \leq \tau,
\end{array}
\right.
\end{align}
with
\begin{align}
f(t)=4(\lambda_{w, 2}-\lambda_{w,0})(t/\tau)^2+\lambda_{w, 0}.
\end{align}
For this experimental protocol, the average of the dissipated availability, $\langle A \rangle = \int^\tau_0 g_{\mu\nu}^{(1)} \dot{\lambda}_\mu \dot{\lambda}_\nu$, and its fluctuation $\overline{\avg{\Delta A^2}}$ given by Eq.~(9) in the main paper are calculated as
\begin{align}
\langle A_{\rm exp}^{\rm od}  \rangle &= \frac{\gamma k_{\rm B} T_h }{4}\int^{t_1}_0 dt\, \frac{\dot{f}(t)^2}{f(t)^3} +
\frac{\gamma k_{\rm B} T_c }{4}\int^{t_3}_{t_2} dt\, \frac{\dot{f}(\tau-t)^2}{f(\tau-t)^3}, \\
\overline{\langle (\Delta A_{\rm exp}^{\rm od})^2  \rangle} &= \frac{\gamma (k_{\rm B}T_h)^2 }{2}\int^{t_1}_0 dt\, \frac{\dot{f}(t)^2}{f(t)^3} +
\frac{\gamma (k_{\rm B}T_c)^2 }{2}\int^{t_3}_{t_2} dt\, \frac{\dot{f}(\tau-t)^2}{f(\tau-t)^3},
\end{align}
where $k_{\rm B}$ is the Boltzmann constant, $t_i$ is the time at node $i$, and we have used Eqs.~(16) and (21) in the main paper. For clarity, we put the superscript ``od'' representing the overdamped approximation since we will compare the results with and without the overdamped approximation in the next section. Note that $A_{\rm exp}^{\rm od}$ here is identical to $A_{\rm exp}$ in the main paper.

On the one hand, $\langle A \rangle$ and $\overline{\avg{\Delta A^2}}$ obtained by the optimized protocol given by Eq.~(25) in the main text are
\begin{align}
\langle A_{\rm opt} \rangle &= \biggl(\frac{1}{t_1}+\frac{1}{t_3-t_2}\biggr)(\mathcal{L}^{(1)}_h)^2, \\
\overline{\langle \Delta A^2_{\rm opt} \rangle} &=  \biggl(\frac{1}{t_1}+\frac{T_c}{T_h}\frac{1}{t_3-t_2}\biggr)(\mathcal{L}^{(2)}_h)^2,
\end{align}
where 
\begin{align}
\mathcal{L}_h^{(i)} =\sqrt{2^{i-1} \gamma\, (k_{\rm B}T_h)^i}\, (\lambda_{w,1}^{-1/2}-\lambda_{w,0}^{-1/2})
\end{align}
with $i=1$ and $2$.

The experimental value of each parameter is the following~\cite{Martinez16suppl}: The stiffness $\lambda_{w,i}$'s are $\lambda_{w,0}=20.0$~pN$\cdot \mu$m$^{-1}$, $\lambda_{w,1}=6.2$~pN$\cdot \mu$m$^{-1}$, and $\lambda_{w,2}=2.0$~pN$\cdot \mu$m$^{-1}$. The cycle period $\tau$ is $200$~ms and the time at each node is $t_{1} = 0.26 \tau$, $t_{2} =  0.5 \tau$, and $t_{3} = 0.75 \tau$ (i.e., the duration of each stroke is $\tau_{0 \rightarrow 1} = 0.26 \tau$, $\tau_{1 \rightarrow 2} = 0.24 \tau$, $\tau_{2 \rightarrow 3} = 0.25 \tau$, and $\tau_{3 \rightarrow 4} = 0.25 \tau$). The temperatures of the isothermal strokes are $T_c=300$~K and $T_h=525$~K. The friction coefficient can be estimated from Stokes' law, $\gamma = 6 \pi \eta r$, where $\eta$ is the dynamic viscosity of water and $r$ is the radius of the Brownian particle. The viscosity at room temperature is $\eta =0.89$~pN$\cdot\mu$m$^{-2}\cdot$ms \cite{Mestres14suppl} and the radius of the Brownian particle used in the experiment is $r=0.5$~$\mu$m. We then get $\gamma=8.4$~pN$\cdot \mu$m$^{-1}\cdot$ms.   

From these parameter values, the ratio between the optimized value and the experimental one is calculated for $\avg{A}$ and $\overline{\avg{\Delta A^2}}$ respectively as 
\begin{align}
\frac{\langle A_{\rm opt} \rangle}{\langle A_{\rm exp}^{\rm od} \rangle}=0.65, \qquad  \frac{\,\,\,\overline{\langle \Delta A^2_{\rm opt} \rangle}\,\,\,}{\overline{\langle (\Delta A_{\rm exp}^{\rm od})^2 \rangle}}=0.70.
\end{align}
The optimized protocol improves the average of the dissipated availability by $35$\% and its fluctuation by $30$\%.

\subsection{Dissipated availability in the isentropic and isothermal processes in the experiment of Ref.~\cite{Martinez16suppl}}

In the overdamped approximation employed in the main paper, the dissipated availability and the thermodynamics length of the isentropic strokes are exactly zero. However, in the actual experiments, the system is not perfectly overdamped, and these quantities are nonzero. Nevertheless, here we show that the dissipated availability of the isentropic strokes is much smaller than that of the isothermal strokes in the experiment of Ref.~\cite{Martinez16suppl} for the cycle period $\tau = 200$~ms considered in the main paper, which validates the overdamped approximation for this case.

For the underdamped Brownian motion described by the following Langevin equation with the Gaussian white noise $\zeta$:
\begin{align}
m \ddot{q} &= -\gamma \dot{q} - \lambda_w(t)\, q+\zeta, \\
\langle \zeta(t)\, \zeta(t')\rangle &= 2 \gamma k_B T(t)\, \delta(t-t'),
\end{align}
the metric $g^{(1)}_{\mu\nu}$ is given by~\cite{Frim21suppl}
\begin{align}
  g^{(1)}_{\mu\nu} = \frac{\gamma}{4k_BT\lambda_w}
  \begin{pmatrix}
  \left(\dfrac{\strut k_BT}{\strut \lambda_w}\right)^2 (1 + \chi) &
    -\dfrac{\strut k_BT}{\strut \lambda_w} (1 + 2\chi) \\
  -\dfrac{\strut k_BT}{\strut \lambda_w} (1 + 2\chi) &
    1 + 4\chi
  \end{pmatrix},\label{eq:g1suppl}
\end{align}
where $\chi \equiv m\lambda_w/\gamma^2$ is a dimensionless parameter characterizing the effect of the inertia and $m$ is the mass of the Brownian particle. The particle used in the experiment of Ref.~\cite{Martinez16suppl} is made of polystyrene with the mass density $\rho \simeq 1.0$~g/cm$^3$. We then estimate the mass of the particle as $m = (4/3) \pi r^3 \rho \simeq 5.2\times 10^{-4} $pN$\cdot\mu$m$^{-1}\cdot$ms$^2$ for $r=0.5$~$\mu$m. Here we can see that, in the overdamped regime where the effect of the inertia is negligible, $\chi \simeq (1.5\times 10^{-5} \mbox{--} 1.5 \times 10^{-4}) \ll 1$, the metric tensor $g^{(1)}_{\mu\nu}$ in the underdamped case [Eq.~(\ref{eq:g1suppl})] reduces to the one in the overdamped approximation given by Eq.~(21) in the main paper together with the relation $g^{(1)}_{\mu\nu} = (2k_BT)^{-1} g^{(2)}_{\mu\nu}$.

With the metric given by Eq.~(\ref{eq:g1suppl}) and the protocols of $T(t)$ and $\lambda_w(t)$ in the experiment of Ref.~\cite{Martinez16suppl} for the underdamped case, we evaluate the average of the dissipated availability $A_{\rm exp}^{\rm ud}(\mbox{isothermal})$ and $A_{\rm exp}^{\rm ud}(\mbox{isentropic})$ in the two isothermal and the two isentropic strokes of the Brownian Carnot cycle (the superscript ``ud'' represents the underdamped case). Note that, unlike the isentropic processes in the overdamped approximation given by $T(t)/\lambda_w(t) = \mbox{const.}$, the isentropic processes here is given by $T^2(t)/\lambda_w(t) = \mbox{const.}$\,.

The average of the dissipated availability in the isothermal and the isentropic strokes are given respectively by
\begin{align}
\langle A_{\rm exp}^{\rm ud}(\mbox{isothermal}) \rangle &=\biggl(\int^{t_1}_0 dt +\int^{t_3}_{t_2} dt\biggr) g_{w w}^{(1)}(t)\, \dot{\lambda}_w(t)\, \dot{\lambda}_w(t), \\
\langle A_{\rm exp}^{\rm ud}(\mbox{isentropic}) \rangle &=\biggl(\int_{t_1}^{t_2} dt +\int_{t_3}^{t_4} dt\biggr) g_{\mu \nu}^{(1)}(t)\, \dot{\lambda}_\mu(t)\, \dot{\lambda}_\nu(t).
\end{align}
For the same time $t_i$ (i.e., $t_1=0.26 \tau$, $t_2 = 0.5 \tau$, and $t_3=0.75 \tau$ with $\tau = 200$~ms), $T(t_i)$, and $\lambda_w(t_i)$ of the four nodes as in the experiment of Ref.~\cite{Martinez16suppl} given in the previous section, we calculate $\avg{A_{\rm exp}^{\rm ud}(\mbox{isothermal}}$ and $\avg{A_{\rm exp}^{\rm ud}(\mbox{isentropic})}$.

First, the result of the ratio between $\avg{A_{\rm exp}^{\rm ud}(\mbox{isentropic}}$ and $\avg{A_{\rm exp}^{\rm ud}(\mbox{isothermal})}$ is
\begin{align}
  \frac{\avg{A_{\rm exp}^{\rm ud}(\mbox{isentropic})}}{\avg{A_{\rm exp}^{\rm ud}(\mbox{isothermal}}} = 0.057.
\end{align}
Thus, in the experiment, the dissipation in the isentropic strokes is much smaller than that in the isothermal strokes and is only $5$\% of the total dissipation in the whole cycle.

Furthermore, the total dissipated availability in the experiment without the overdamped approximation, $A_{\rm exp}^{\rm ud} \equiv A_{\rm exp}^{\rm ud}(\mbox{isothermal}) + A_{\rm exp}^{\rm ud}(\mbox{isentropic})$, and that within the overdamped approximation, $A_{\rm exp}^{\rm od}$ (which is identical to $A_{\rm exp}$ in the main paper), are almost the same. The relative difference between the average of these quantities is
\begin{align}
  \frac{\avg{A_{\rm exp}^{\rm ud}} -  \avg{A_{\rm exp}^{\rm od}}}{\avg{A_{\rm exp}^{\rm ud}}} = 0.054.
\end{align}
Since the dissipation $A_{\rm exp}^{\rm od}$ within the overdamped approximation solely comes from the isothermal strokes, it is also worthwhile to compare $A_{\rm exp}^{\rm od}$ and $A_{\rm exp}^{\rm ud}(\mbox{isothermal})$:
\begin{align}
  \frac{\avg{A_{\rm exp}^{\rm od}}}{\avg{A_{\rm exp}^{\rm ud}(\mbox{isothermal})}} = 1.00004.
\end{align}
Namely, the difference between the experiment without the overdamped approximation and the overdamped case is negligible.

In conclusion, in the experiment of Ref.~\cite{Martinez16suppl} (for $\tau = 200$ ms) considered in the main paper, dissipation $\avg{A_{\rm exp}^{\rm ud}(\mbox{isentropic})}$ in the isentropic strokes is negligible, and the total dissipation $\avg{A_{\rm exp}^{\rm ud}}$ can be well evaluated by the overdamped approximation $\avg{A_{\rm exp}^{\rm od}}$ (which is identical to $\avg{A_{\rm exp}}$ in the main paper) around $5$\% error. This analysis shows that the overdamped approximation is valid and gives good estimates for the experiment of Ref.~\cite{Martinez16suppl}.


\begin{thebibliography}{99}
\bibitem{note:isentropic} In the experiment of Ref.~\cite{Martinez16}, adiabatic strokes in the conventional Carnot cycle are replaced by isentropic strokes with a constant Shannon entropy of the working substance. In the present overdamped case, such an isentropic process is given by $T/\lambda_w = \mbox{const.}$ instead of $T^2/\lambda_w = \mbox{const.}$ for the underdamped case.
\bibitem{note:fluct} It is noted that, although a separate isentropic stroke gives nonzero fluctuation of $A$ due to the contribution from its end points, the isentropic strokes in a cycle make no contribution to $\overline{\avg{\Delta A^2}}$.
\bibitem{note:trivialstrokes} Here we exclude the cycles with trivial strokes in which the system moves along the same path in the parameter space back and forth.
\end{thebibliography}
\end{document}